\documentclass[letter,onecolumn]{jpsj2}

\def\runauthor{Kazutaka \textsc{Kudo} {\it et al.}}

\title{%
Drastic Enhancement of Thermal Conductivity in the Bose-Einstein Condensed State of TlCuCl$_3$
}

\author{Kazutaka \textsc{Kudo}\thanks{E-mail address: kudo@imr.tohoku.ac.jp}\thanks{Present address: Institute for Materials Research, Tohoku University, Katahira 2-1-1, Aoba-ku, Sendai 980-8577}, 
Mitsuhiro \textsc{Yamazaki}, Takayuki \textsc{Kawamata}, 
Takashi \textsc{Noji}, Yoji \textsc{Koike}, Terukazu \textsc{Nishizaki}$^1$, Norio \textsc{Kobayashi}$^1$ and Hidekazu \textsc{Tanaka}$^2$
}

\inst{%
Department of Applied Physics, Graduate School of Engineering, Tohoku University, Aoba-yama 05, Aoba-ku, Sendai 980-8579 \\ 
$^1$Institute for Materials Research, Tohoku University, Katahira 2-1-1, Aoba-ku, Sendai 980-8577 \\
$^2$Research Center for Low Temperature Physics, Tokyo Institute of Technology, Oh-okayama, Meguro-ku, Tokyo 152-8551
}

\recdate{}

\abst{%
We have measured the thermal conductivity of a TlCuCl$_3$ single crystal in magnetic fields up to 14 T. 
It has been found that the temperature dependence of the thermal conductivity exhibits a sharp peak at 4 K in zero field, which is suppressed by the application of magnetic fields up to 7 T.
The peak is concluded to be attributable to the enhancement of the thermal conductivity due to phonons because of the formation of a spin-gap state. 
In high magnetic fields above 7 T, on the other hand, another sharp peak appears around 4 K and this is enhanced with increasing magnetic field. 
This peak is regarded as being attributable to the enhancement of the thermal conductivity due to magnons and/or phonons because of the drastic extension of the mean free path of magnons  and/or phonons in the Bose-Einstein condensed state.
}

\kword{%
Bose-Einstein condensation, thermal conductivity, magnon, phonon, spin gap, dimer
}

\begin{document}
\sloppy
\maketitle

Recently, the thermal conductivity in quantum spin systems has attracted considerable interest because it is closely related to the spin state.
For example, in a spin-gap state, both the phonon-magnon scattering rate  and the magnon-magnon scattering rate markedly decrease, so that both $\kappa_{\rm phonon}$ (thermal conductivity due to phonons) and $\kappa_{\rm magnon}$ (thermal conductivity due to magnons) are expected to be enhanced. 
Actually, enhancement of $\kappa_{\rm phonon}$ has been observed in the spin-gap state of the spin-Peierls system CuGeO$_3$\cite{rf:CuGeO} and also of the two-dimensional orthogonal spin-dimer system SrCu$_2$(BO$_3$)$_2$\cite{rf:KudoSrCu2,rf:HoffSrCu2}.
As for the enhancement of $\kappa_{\rm magnon}$, on the other hand, it depends on the magnitude of the bandwidth of the excited state of magnons whether $\kappa_{\rm magnon}$ is enhanced or suppressed. 
That is, in systems with a large bandwidth such as the two-leg spin-ladder system Sr$_{14}$Cu$_{24}$O$_{41}$\cite{rf:KudoLT,rf:Solo,rf:Kudo1,rf:Kudo2,rf:Hess}, the enhancement has been observed, while the suppression has been observed in systems with a small bandwidth such as CuGeO$_3$\cite{rf:CuGeO}. 
In the antiferromagnetically ordered state of the two-dimensional spin system Cu$_3$B$_2$O$_6$, meanwhile, a marked increase in $\kappa_{\rm magnon}$ has been observed at low temperatures below the N\'{e}el temperature $T_{\rm N}$\cite{rf:kudo326-2}.
Furthermore, it has been found that the thermal conductivity is sensitive to changes of the spin state in magnetic fields, such as reduction of a spin-gap and a spin-flop transition\cite{rf:CuGeO,rf:KudoSrCu2,rf:HoffSrCu2,rf:kudo326-2}.
Accordingly, it is considered that the thermal conductivity measurement is a potential probe detecting a change of the spin state as a function of temperature and/or magnetic field.

\begin{figure}[tb]
\begin{center}
\includegraphics[width=1\linewidth]{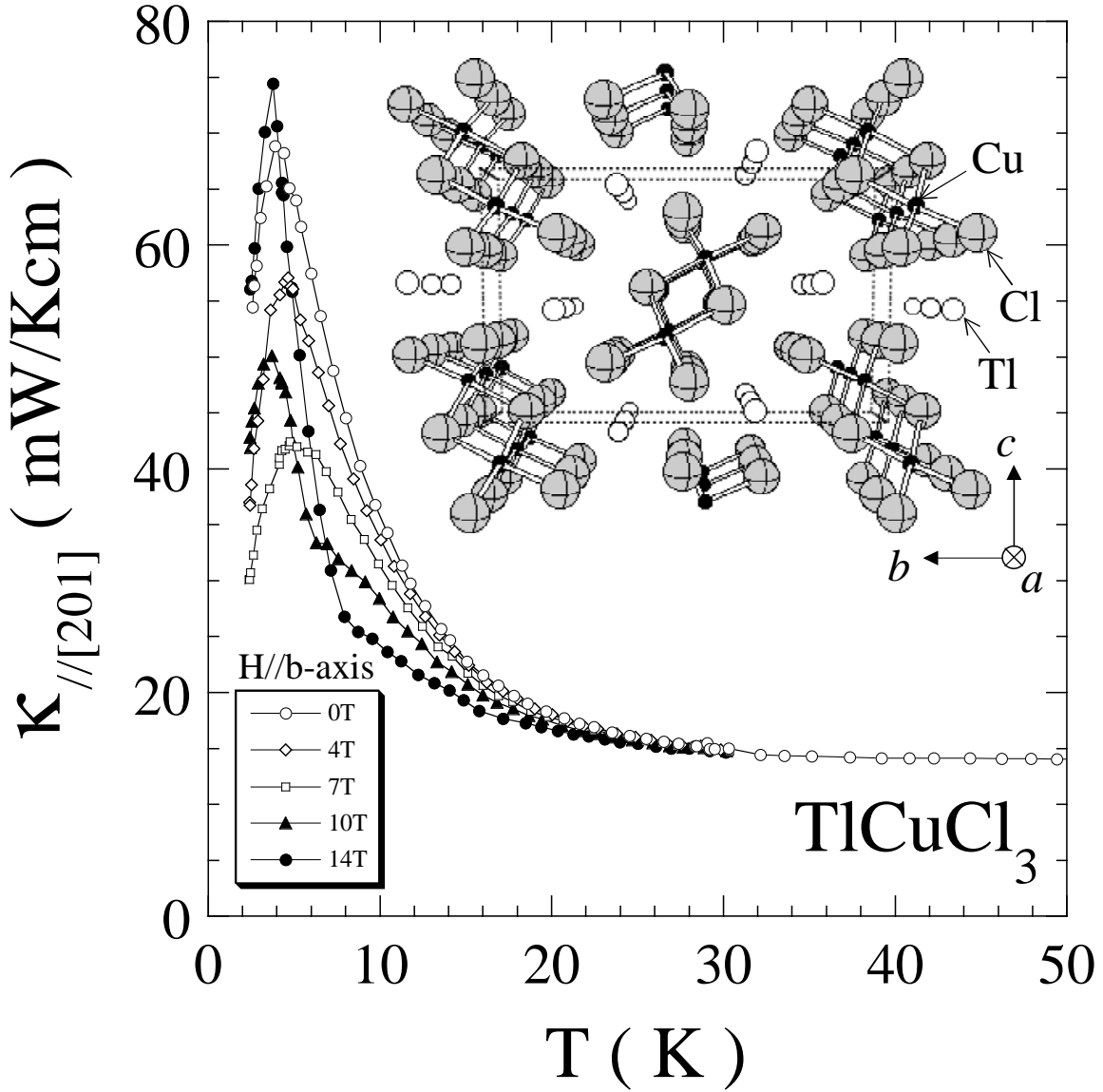}
\end{center}
\caption{
Temperature dependence of the thermal conductivity along the [2 0 1] direction in magnetic fields parallel to the $b$-axis.
The inset shows the crystal structure of TlCuCl$_3$. The dotted parallelogram indicates the unit cell.
 }\label{fig:1}
 \end{figure}

In this paper, we have applied the thermal conductivity measurement to a quantum spin system TlCuCl$_3$, which exhibits changes of the spin state, depending on temperature and magnetic field.
The crystal structure of this material is monoclinic and composed of planar dimers of Cu$_2$Cl$_6$\cite{rf:Takatsu}, as shown in the inset of Fig. \ref{fig:1}.
The dimers are stacked on top of one another and form infinite double chains parallel to the $a$-axis.
These double chains are located at the corners and center of the unit cell in the $bc$-plane and are separated by Tl$^+$ ions. 
As for the spin state of TlCuCl$_3$, the ground state is a spin-gap state with the excitation gap $\Delta =$ 7.7 K in zero field\cite{rf:Takatsu,rf:Shiramura,rf:Oosawa}.
From the analysis of the dispersion relation obtained by the inelastic neutron scattering experiment, it has been found that the spin gap of TlCuCl$_3$ is due  to the strong antiferromagnetic interaction $J =$ 5.68 meV in the planar dimer of Cu$_2$Cl$_6$ and that the neighboring dimers are coupled by the strong interdimer interactions along the double chain and in the (1 0 $\bar{2}$) plane \cite{rf:Cavadini,rf:Oosawa2}.
Our motivation of the thermal conductivity measurement for TlCuCl$_3$ arises from the three-dimensional magnetic ordering induced by the application of magnetic fields higher than the gap field $H_{\rm g} \sim$ 5.5 T\cite{rf:Oosawa}.
Under a constant magnetic field above $\sim$ 5.5 T, the temperature dependence of the magnetization exhibits a cusplike minimum at $T_{\rm N}$, and the ordering temperature increases with increasing magnetic field\cite{rf:Shiramura}. 
From a theoretical point of view, both the observed temperature dependence of the magnetization and the field dependence of the ordering temperature have been qualitatively well described in terms of the Bose-Einstein condensation (BEC) of magnons\cite{rf:Nikuni}.
Actually, this suggestion has been confirmed by the observation of ordering of the transverse component of spins in the elastic neutron scattering experiment\cite{rf:Tanakaneutron}. 
Moreover, very recently, the Goldstone mode characteristic of the BEC of magnons, theoretically suggested by Matsumoto {\it et al.}\cite{rf:Matsumoto}, has actually been observed in the inelastic neutron scattering experiment by R\"{u}egg {\it et al.}\cite{rf:Ruegg}.
In the BEC state, the behavior of the thermal conductivity is an attractive issue, because very large values of $\kappa_{\rm magnon}$ are expected from the analogy of the thermal conductivity in the superfluid (BEC) state of liquid $^4$He\cite{rf:17}.
Therefore, in this study, we measured the thermal conductivity of TlCuCl$_3$ along the [2 0 1] direction, where the interdimer coupling is the strongest, and investigated the temperature and magnetic-field dependences.
As a result, we observed drastic enhancement of the thermal conductivity in the BEC state for the first time.

Single crystals of TlCuCl$_3$ were grown by the vertical Bridgman method. 
The details are described in the previous paper\cite{rf:Oosawa}.
Thermal conductivity measurements were carried out by a conventional steady-state method. 
One side of a rectangular single crystal (7.0 $\parallel$ [2 0 1] $\times$ 1.0 $\times$ 1.0 mm$^3$) was anchored on the copper heat sink with indium solder. 
A chip resistance of 1 k$\Omega$ was attached as a heater to the opposite side of the single crystal with GE7031 varnish. 
The temperature difference across the crystal (0.01--1.0 K) was measured with two Cernox thermometers (LakeShore Cryotronics, Inc., CX-1050-SD). 
The system was proved to be satisfactory from the thermal conductivity measurement of a standard reference material, Austenitic Stainless Steel. 
Magnetic fields up to 14 T were applied parallel to the $b$-axis\cite{mag}.

\begin{figure}[tb]
\begin{center}
\includegraphics[width=1\linewidth]{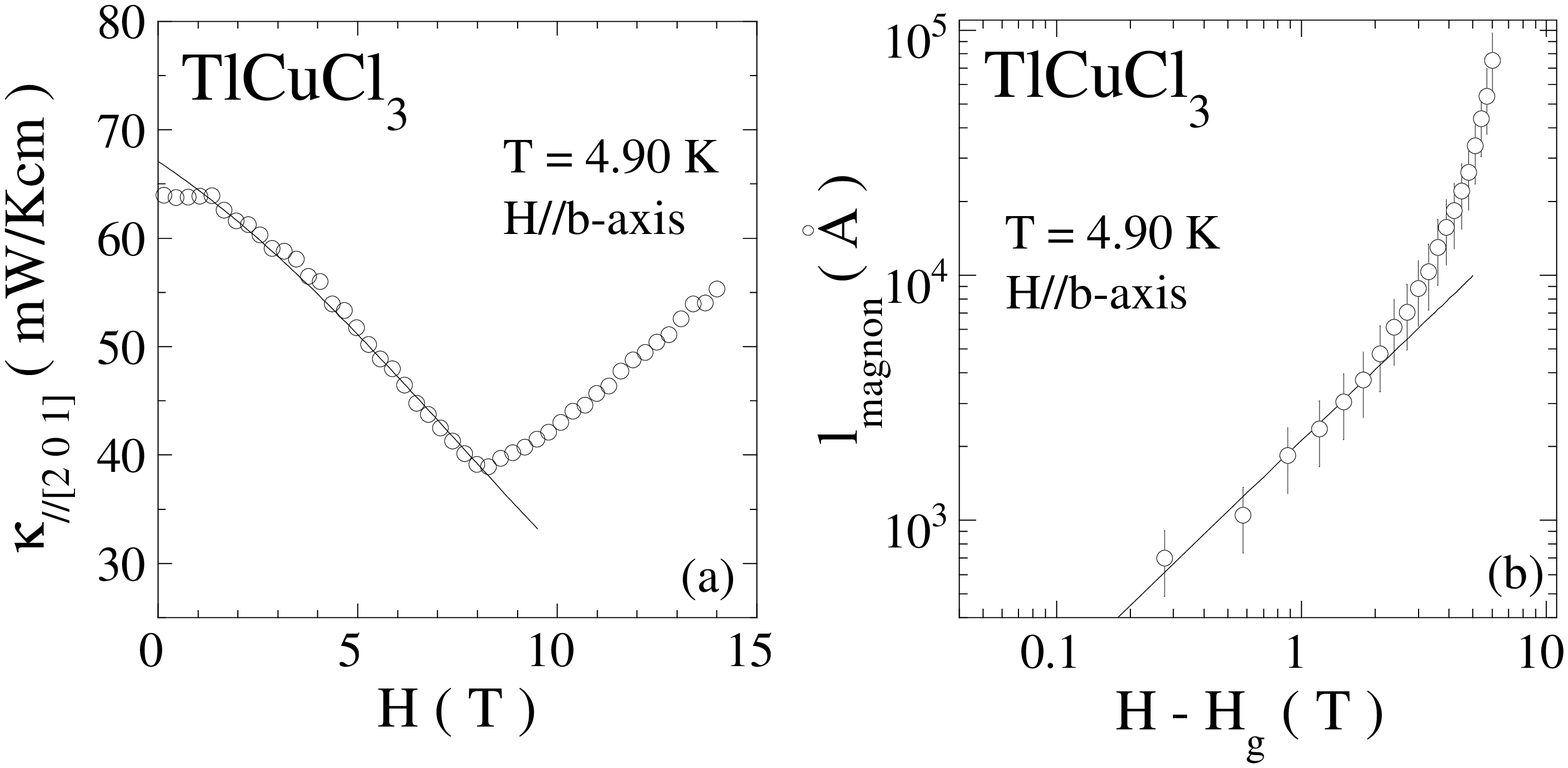}
\end{center}
\caption{
(a) Magnetic-field dependence of the thermal conductivity along the [2 0 1] direction at 4.90 K. 
Magnetic fields were applied along the $b$-axis. 
The solid line denotes the best-fit curve of 1/\{$a + b{\rm exp}(-\frac{\Delta(H)}{T})$\}. 
(b) Magnetic-field dependence of the mean free path of magnons along the [2 0 1] direction at 4.90 K. 
Error bars are due to errors of $C_{\rm magnon}$\cite{rf:C}. 
The solid line denotes a fitting curve proportional to $|H - H_{\rm g}|^n$ with $n =$ 1.0. 
}
\label{fig:2}
\end{figure}

Figure \ref{fig:1} shows the temperature dependence of the thermal conductivity along the [2 0 1] direction in magnetic fields parallel to the $b$-axis up to 14 T.
It is found that the thermal conductivity in zero field increases with decreasing temperature at low temperatures below 40 K and shows a sharp peak at 4 K. 
With the application of magnetic fields up to 7 T, the peak is suppressed. 
In high magnetic fields above 7 T, on the other hand, the thermal conductivity increases with decreasing temperature and exhibits a kink and another sharp peak around 4 K.
This unusual field dependence of the thermal conductivity is clearly seen in Fig. \ref{fig:2}(a). 

First, we discuss the origin of the sharp peak observed around 4 K in low magnetic fields below 7 T. 
Taking into account the theoretical calculation by Matsumoto\cite{rf:MatsumotoP} that the velocity of the excited state of magnons, namely, triplet magnons, is negligibly small in the spin-gap state of TlCuCl$_3$, the contribution of $\kappa_{\rm magnon}$ is neglected. 
Hence, at a glance, the temperature dependence of the thermal conductivity in zero field appears to be regarded as a typical behavior of $\kappa_{\rm phonon}$ which exhibits a peak at a low temperature as a consequence of the freezing out of Umklapp processes of phonons with decreasing temperature on the one hand and the decreasing number of phonons on the other. 
However, the case is not simple, because Umklapp processes are generally independent of magnetic field so that the suppression of the peak in magnetic fields cannot be explained simply. 
It is likely that the peak is related to the formation of the spin-gap state, because the increase in the thermal conductivity with decreasing temperature below 40 K is in good correspondence with the decrease in the magnetic susceptibility below 40 K due to the formation of the spin-gap state\cite{rf:Oosawa}. 
As reported for CuGeO$_3$\cite{rf:CuGeO} and SrCu$_2$(BO$_3$)$_2$\cite{rf:KudoSrCu2,rf:HoffSrCu2}, in fact, $\kappa_{\rm phonon}$ is enhanced in the spin-gap state, owing to the reduction of the phonon-magnon scattering. 
The peak is regarded as a result of the formation of the spin gap on the one hand and the decreasing number of phonons on the other. 
In magnetic fields, on the other hand, the suppression of the peak is interpreted as being attributable to the enhancement of the phonon-magnon scattering because of the reduction of the spin gap. 
Actually, the magnetic-field dependence of the peak in magnetic fields up to 7 T is similar to those in the spin-gap state of CuGeO$_3$\cite{rf:CuGeO} and SrCu$_2$(BO$_3$)$_2$\cite{rf:KudoSrCu2,rf:HoffSrCu2}.

To prove this interpretation, we perform a fitting analysis for the magnetic-field dependence of the thermal conductivity at 4.90 K shown in Fig. \ref{fig:2}(a). 
Usually, $\kappa_{\rm phonon}$ is given by $\kappa_{\rm phonon} = \frac{1}{3}C_{\rm phonon}v_{\rm phonon}^2\tau_{\rm phonon}$, where $C_{\rm phonon}$ is the specific heat of phonons, $v_{\rm phonon}$ the velocity of phonons and $\tau_{\rm phonon}$ the relaxation time of phonons. 
The scattering rate of phonons $\tau_{\rm phonon}^{-1}$ is described as $\tau_{\rm phonon}^{-1} = \sum_i \tau_i^{-1}$, with $i$ denoting different scattering processes such as the phonon-magnon, phonon-boundary, phonon-defect and Umklapp scatterings. 
Changes of $C_{\rm phonon}$ and $v_{\rm phonon}$ by the application of magnetic field are usually as small as at most 1\%, even if softening of phonons occurs because of the change of the spin state\cite{rf:CuGeO,rf:KudoSrCu2,rf:HoffSrCu2,rf:Saint,rf:vs}. 
Therefore, the magnetic-field dependences of $C_{\rm phonon}$ and $v_{\rm phonon}$ are neglected. 
In this simple analysis, only the magnetic-field dependence of $\tau_{\rm phonon}$ due to the phonon-magnon scattering is taken into account. 
The magnetic-field dependence of $\tau_{\rm phonon}^{-1}$ is assumed to be simply due to the number of triplet magnons, so that the magnetic-field dependence of $\kappa_{\rm phonon}$ is given by $\kappa_{\rm phonon}(H) \sim 1/\{a + b {\rm exp}({-\frac{\Delta(H)}{T}})\}$, where $a$ and $b$ are constants and $\Delta(H)$ is the spin gap changing linearly with magnetic field\cite{rf:Oosawa,rf:Cavadini, rf:Matsumoto}. 
As shown in Fig. \ref{fig:2}(a), the best fit curve is well overlapped on the experimental data points in low magnetic fields below 8 T. 
Thus, the suppression of the thermal conductivity with increasing magnetic field below 8 T at 4.90 K is well described in terms of the increase in the phonon-magnon scattering rate. 
Accordingly, it is proved that the sharp peak observed around 4 K in low magnetic fields below 7 T is due to the enhancement of $\kappa_{\rm phonon}$ in the spin-gap state and that the suppression of the peak in low magnetic fields is attributed to the increase in the phonon-magnon scattering rate because of the reduction of the spin gap.

\begin{figure}[tb]
\begin{center}
\includegraphics[width=0.9\linewidth]{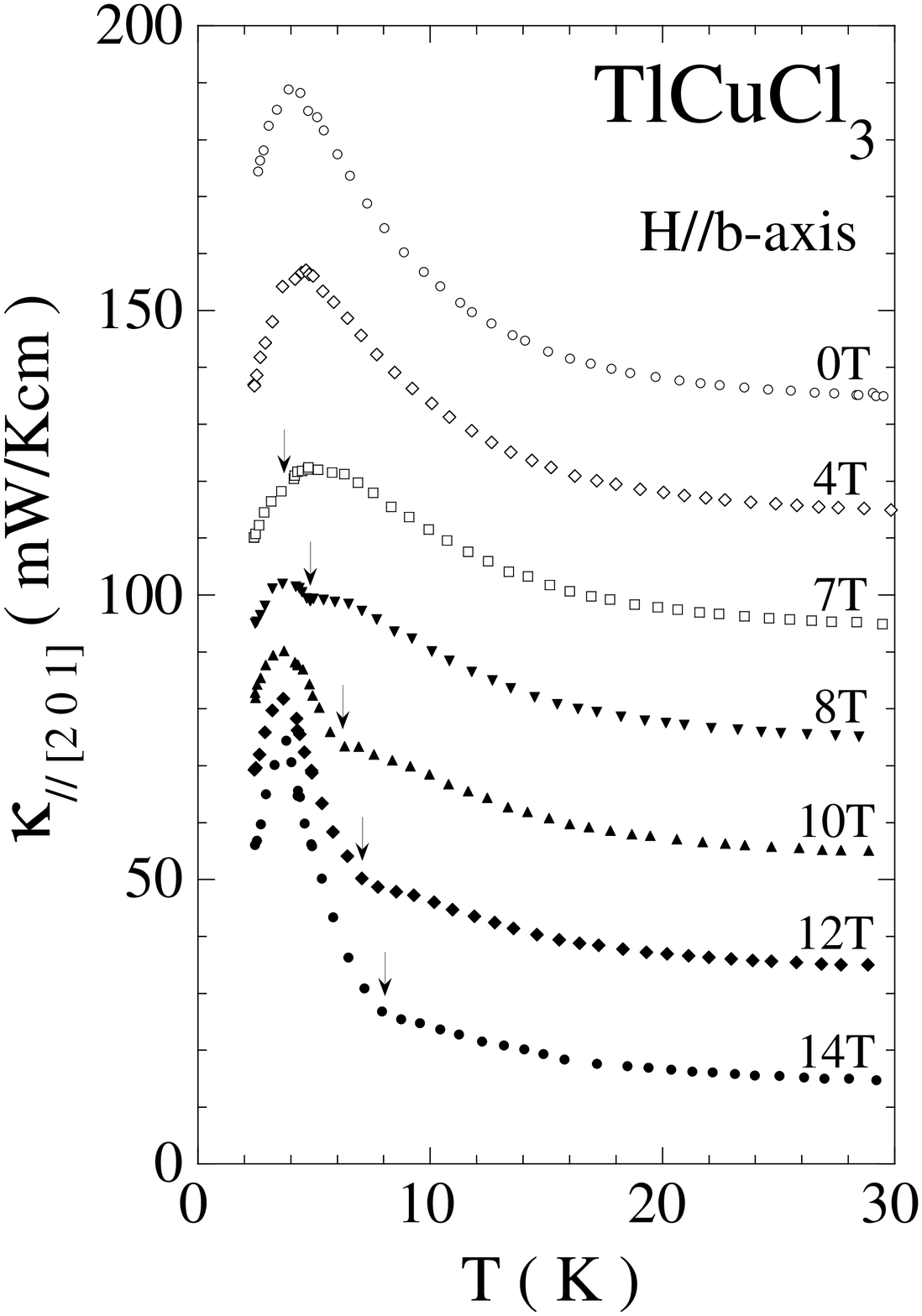}
\end{center}
\caption{
Low-temperature part of the temperature dependence of the thermal conductivity shown in Fig. \ref{fig:1}. 
Each plot in a magnetic field is shifted vertically by 20 mW/Kcm for clarity. 
Arrows indicate the temperature below which another peak develops with decreasing temperature.
}\label{fig:3}
 \end{figure}
\begin{figure}[tb]
 \begin{center}
\includegraphics[width=0.9\linewidth]{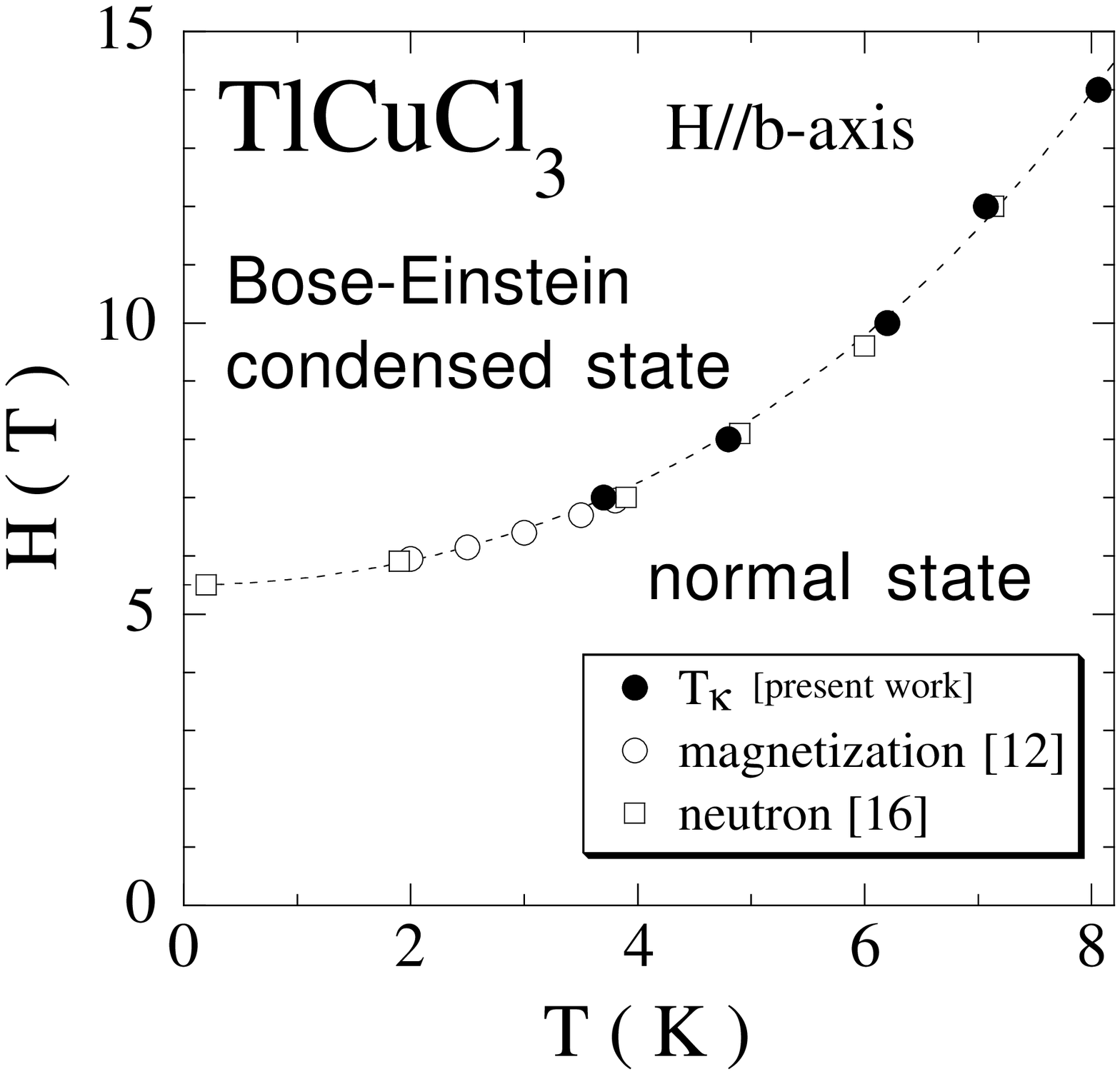}
\end{center}
\caption{Phase boundary between the Bose-Einstein condensed state and the normal state in TlCuCl$_3$, determined from the thermal conductivity, magnetization\cite{rf:Oosawa} and neutron scattering\cite{rf:Tanakaneutron} measurements. 
Magnetic fields were applied parallel to the $b$-axis. The dashed line is a guide for the eyes.}
\label{fig:4}
\end{figure}

Next, we discuss the origin of another sharp peak observed around 4 K in high magnetic fields above 7 T.
Here, we define $T_\kappa$ as the temperature where the thermal conductivity exhibits a kink, as indicated by arrows in Fig. \ref{fig:3}.
It is found that $T_\kappa$ depends on the magnetic field $H$. 
$T_\kappa$ values are clearly determined as the local minimum points in the $\kappa(H) - \kappa(0)$ vs $T$ plot and plotted in the $H-T$ phase diagram in Fig. \ref{fig:4} together with BEC temperatures estimated from the magnetization\cite{rf:Oosawa} and the elastic neutron scattering\cite{rf:Tanakaneutron} measurements. 
It is found that $T_\kappa$ is in good agreement with the BEC temperature.
Furthermore, the inflection field (8 T at 4.90 K) in the magnetic-field dependence of the thermal conductivity shown in Fig. \ref{fig:2}(a) is located on the phase boundary between the BEC state and the normal state, and hereby is regarded as $H_{\rm g}$ at 4.90 K. 
Therefore, it is concluded that the origin of the sharp peak around 4 K in high magnetic fields above 7 T is due to BEC.

Finally, we discuss whether the sharp peak around 4 K in high magnetic fields above 7 T is due to the enhancement of $\kappa_{\rm phonon}$ or $\kappa_{\rm magnon}$. 
As mentioned above, $\kappa_{\rm magnon}$ is expected to bring about the enhancement on the basis of the analogy of the thermal conductivity in the superfluid state of liquid $^4$He being much larger than that in the normalfluid state\cite{rf:17}. 
On the basis of the two-fluid model\cite{rf:Hetext}, it is understood that the mean free path of the remaining normalfluid 
carrying heat becomes drastically long in the superfluid state, because the normalfluid-superfluid scattering rate is very small.
In the case of TlCuCl$_3$, by analogy, the mean free path of the normalfluid magnons may be expected to become drastically long in the BEC state so that $\kappa_{\rm magnon}$ is drastically enhanced. 
At much lower temperatures below the BEC temperature, on the other hand, $\kappa_{\rm magnon}$ should decrease with decreasing temperature, owing to the decreasing number of normalfluid magnons. 
This analogy may be plausible, but it is not sure whether such a hydrodynamic heat conduction due to the mutual conversion of the normalfluid and superfluid in liquid $^4$He is valid in TlCuCl$_3$ as well. 
However, it is worthwhile examining the extension of the mean free path of magnons, $l_{\rm magnon}$, in the BEC state of magnons. 
In the following, we estimate the magnetic-field dependence of $l_{\rm magnon}$ from the magnetic-field dependence of the thermal conductivity at 4.90 K shown in Fig. \ref{fig:2}(a). 
$\kappa_{\rm magnon}(H)$ is expressed as $\kappa_{\rm magnon}(H) = C_{\rm magnon}(H)v_{\rm magnon}(H)l_{\rm magnon}(H)$, with $C_{\rm magnon}(H)$ being the specific heat of magnons and $v_{\rm magnon}(H)$ the velocity of magnons. 
Assuming that the thermal conductivity at $H_{\rm g} =$ 8 T is equal to $\kappa_{\rm phonon}$ and that $\kappa_{\rm phonon}$ is independent of magnetic field above $H_{\rm g}$, $\kappa_{\rm magnon}(H)$ is approximately estimated to be $\kappa(H) - \kappa(H_{\rm g})$. 
On the other hand, $C_{\rm magnon}(H)$ is approximately estimated to be $C(H) - C(0)$ from the specific heat measurements by Oosawa {\it et al.}\cite{rf:C}, and $v_{\rm magnon}(H)$ has been theoretically calculated by Matsumoto\cite{rf:MatsumotoP}. 
The values of calculated $\l_{\rm magnon}(H)$ are thus shown in Fig. \ref{fig:2}(b). 
Error bars are due to errors of $C_{\rm magnon}(H)$, because $C(0)$ includes the Shottky-type specific heat of magnons\cite{rf:C}. 
It is actually found that $l_{\rm magnon}$ exhibits a rapid increase with increasing magnetic field above $H_{\rm g}$. 
$l_{\rm magnon}$ increases in proportion to $|H - H_{\rm g}|^n$ with $n = 1.0$ around $H_{\rm g}$ and is as large as 7.5 $\times$ 10$^4$ ${\rm \AA}$ at 14 T. 
Above $H_{\rm g}$, $v_{\rm magnon}$ also increases with increasing magnetic field, but the value at 14 T is at most 2.0 $\times$ 10$^5$ cm/s\cite{rf:MatsumotoP} which is not so different from the typical value of $v_{\rm phonon}$\cite{rf:Saint,rf:vs}.
On the other hand, the estimated value of $l_{\rm magnon}$ at 14 T is about ten times larger than those of one-dimensional spin systems CuGeO$_3$\cite{rf:Takeya}, SrCuO$_2$\cite{rf:Solo2} and Sr$_2$CuO$_3$\cite{rf:Solo2} where the large contribution of $\kappa_{\rm magnon}$ has been pointed out. 
Such a large value of $l_{\rm magnon}$ in TlCuCl$_3$ may be characteristic of the BEC state of magnons, suggesting the drastic suppression of the magnon-magnon scattering rate as discussed on the basis of the analogy of the superfluid state of liquid $^4$He.
Accordingly, it is likely that the observed sharp peak of the thermal conductivity around 4 K in high magnetic fields above 7 T is due to the drastic enhancement of $\kappa_{\rm magnon}$, owing to the increase in $v_{\rm magnon}$ and the drastic extension of $l_{\rm magnon}$.

On the other hand, it is possible that the sharp peak around 4 K in high magnetic fields above 7 T is due to the enhancement of $\kappa_{\rm phonon}$. 
In this case, the origin of the peak is attributed to the extension of the mean free path of phonons in the BEC state where the phonon-magnon scattering is suppressed as the number of the normalfluid magnons decreases. 
One may claim that $\kappa_{\rm phonon}$ in the BEC state cannot be larger than $\kappa_{\rm phonon}$ in zero field around 4 K, because the phonon-magnon scattering rate in the spin-gap state seems to be always smaller than that in the gapless state such as in the present BEC state. 
However, this may be misleading, because the phonon-magnon scattering in zero field may not be suppressed completely around the peak temperature (4 K) which is comparable with $\Delta =$ 7.7 K. 
To be more conclusive, thermal conductivity measurements for the Mg-substituted TlCuCl$_3$ are under way.

In conclusion, for a TlCuCl$_3$ single crystal, we have measured the thermal conductivity along the [2 0 1] direction in magnetic fields parallel to the $b$-axis up to 14 T.
It has been found that the thermal conductivity is drastically enhanced with increasing magnetic field in the BEC state. 
The enhancement has been regarded as being attributable to the enhancement of the thermal conductivity due to magnons and/or phonons because of the drastic extension of the mean free path of magnons and/or phonons. 
To the best of our knowledge, the present work is the first report on the enhancement of the thermal conductivity in the BEC state of magnons.

\section*{Acknowledgements}
We thank Dr. M. Matsumoto for informing us of their unpublished theoretical results. 
We are indebted to Dr. J. Takeya for useful discussions.
We are also grateful to Professor T. Sasaki for the support in the thermal conductivity measurements, which were performed at the High Field Laboratory for Superconducting Materials, IMR, Tohoku University.
This work was supported by a Grant-in-Aid for Scientific Research from the Ministry of Education, Culture, Sports, Science and Technology, Japan, by a Grant-in-Aid from the Kazuchika Okura Memorial Foundation, by the Iwatani Naoji Foundation's Research Grant and also by CREST of the Japan Science and Technology Corporation. 
One of the authors (K. K.) was supported by the Japan Society for the Promotion of Science.


\begin{thebibliography}{99}
\bibitem{mag} It has been shown by Lake Shore Cryotronics Inc. that the magnetic-field-induced temperature reading error of Cernox thermometers (CX-1050-SD) in 14 T is about 1\% at 2.6 K and less than 1\% at higher temperatures. 
Therefore, the error of the thermal conductivity is estimated to be as small as less than 1\%.
\end{thebibliography}
\end{document}